# UPGRADE OF THE PHOTON BEAMLINE CONTROL SYSTEM ON THE SRS

B.G.Martlew, B.Corker, G.Cox, P.W.Heath, M.T.Heron, A.Oates, W.R.Rawlinson, C.D.Sharp, CLRC Daresbury Laboratory, Warrington WA4 4AD, UK

## Abstract

The SRS is a 2GeV synchrotron light source with 14 beamlines serving approximately 34 experimental stations. Control of the major elements of the beamlines (vacuum pumps, gauges, valves and radiation stops) is the responsibility of the main SRS Control System. As part of the long-term upgrade plan for the SRS Control System a large programme of work has been undertaken to modernize beamline control. This work included: development of Linux based PC front end computers to interface to the existing CAMAC I/O system, replacement of the user interface by graphical synoptic diagrams running on Windows NT PCs, development of an ActiveX control for parameter display/control and a cache server to reduce loading on the rest of the control system. This paper describes the major components of the project; the techniques used to manage the new PCs and discusses some of the problems encountered during development.

## 1 INTRODUCTION

One of the last stages in upgrading the SRS control system to a modern, distributed, client-server design has been the replacement of the experimental beamline control system [1,2]. Previously, this consisted of a single 32-bit minicomputer interfaced to the plant via a serial CAMAC highway and 4 CAMAC crates. The user interface was a simple character based application running on the central minicomputer with remote terminals connected through serial RS232 lines. This arrangement provided a functional and flexible solution but the user interface in particular was difficult to use and understand.

In the upgraded system a Front End Computer (FEC) directly controls each CAMAC crate through a Hytec 1330 PC/CAMAC interface[1]. The FECs are industrial PCs running Linux. They are attached to the Control System network and handle monitoring and control requests from remote clients.

Each experimental station has been provided with a standard desktop PC running Windows NT. These provide the user interface through graphical applications that directly show the physical layout of the beamline together with real-time analogue and status information.

## 2 NEW DEVELOPMENTS

This system was based on a solution already in use on several of the newer experimental stations [3]. However, several changes were needed to allow easier migration from the old control system. Similarly, several improvements and technical developments were introduced during the project and new security and system management issues had to be addressed.

### 2.1 User Interface PCs

Previously, the beamline users had a simple text-based interface to the control system. This was provided through dedicated RS232 serial lines and was coordinated by software running on a centralized mini-computer. It required the user to have a detailed understanding of the beamline and had little visual feedback to alert the user to potential problems on the beamline.

Each experimental station already had some form of data acquisition computer so the possibility of using this for control purposes was considered. Unfortunately, there is little standardization of these systems across the experimental stations and this alternative would have involved developing and supporting several different solutions. Instead it was decided to provide a dedicated desktop PC running Windows NT Version 4.0 for each station.

### 2.2 Linux Front End Systems

The client-server protocol currently used on the SRS runs over UDP/IP and has already been implemented on MS-DOS, OS-9 and Windows NT. The main server program, rpcserv, was ported to Linux and device support for both Borer and Hytec CAMAC stepper motor modules was added. Analogue input and output device support together with support for the status control and interlock monitoring system already existed. The support software for some devices requires precise delays of the order of a few 10s of microseconds to be generated in software. Initially problems were encountered due to the multi-tasking behaviour of Linux. However, careful coding and use of the round-robin scheduling algorithm eventually solved this problem.

RedHat Linux 6.2 provides a very stable and versatile platform for a FEC. At the time of writing some systems have been operational for over 6 months with no unscheduled re-boots.

---

[1] http://www.hytec-electronics.co.uk/1330.html

## 2.3 ActiveX Control

Early in the project an ActiveX control was developed to simplify the representation of control devices in graphical user interface software. This control was written using Visual Basic 6 and is capable of handling communication with the device, colour-coded display of status, formatted display of analogue values, interlock warning messages and status control (Figure 1). This control has also proved to be useful in other projects and has also been used by the Accelerator Physics group.

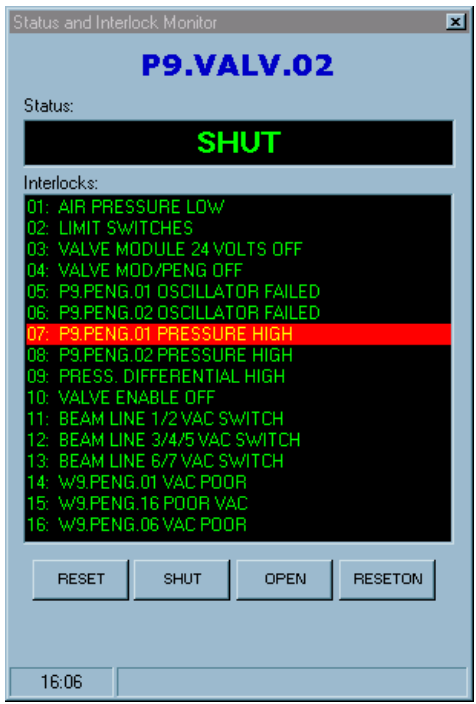

Figure 1: ActiveX control dialog box

## 2.4 Synoptic Control Applications

A graphical synoptic diagram showing the layout of relevant beamline components provides real-time read-back and active control for each experimental station.

These applications were developed using Visual Basic 6 and the ActiveX control described in 2.2 above. As well as the status of the local beamline components the application displays the state of the associated beam port and useful storage ring beam parameters (Figure 2). The displays closely mimic the vacuum flow drawings for the beamline and are produced in co-operation with a beamline control engineer. Online help is available from the application. A contract programmer was hired to do the routine work of implementing and testing the applications.

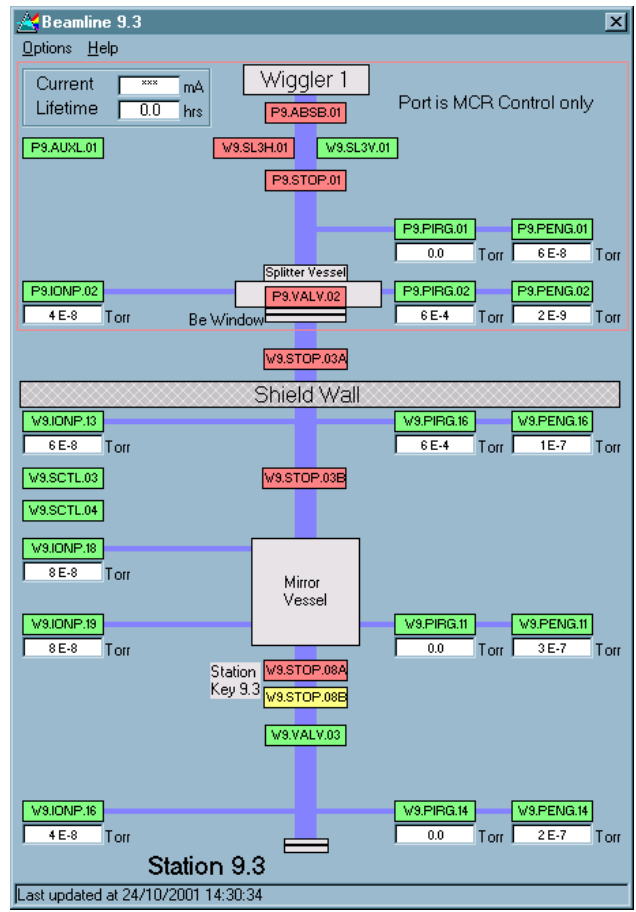

Figure 2: A typical synoptic application

## 2.5 Network upgrade

The SRS control system utilizes 2 class-C networks, one for high-reliability FECs and one for general server and console systems. Fibre optic and twisted-pair cabling had to be installed throughout the experimental area to provide suitable access points for the user interface PCs and Linux FECs.

## 2.6 Cache Server

Initial tests showed that the load on the main control system from 30 or more applications, each requesting port and beam parameters every couple of seconds, was reaching unacceptable levels. To overcome this problem a cache server was introduced. This server was prototyped in Visual Basic using COM objects and subsequently key components have been coded in C++ to provide extra performance and stability. The cache effectively reduces the load on the control system by performing a single read per parameter and serving the result back to each application. The refresh rate is configurable but is typically set at 10 seconds.

## 2.7 Security

Security is provided by limiting user access to the front end computers and by controlling the software allowed to run on the NT workstation.

User access is limited using the NT workstation local policy. A unique user ID is assigned to each experimental station. Only that user ID or a system administrator is allowed to log on locally. The unique ID logs on automatically when NT is started. No network access to the workstation is allowed. The floppy disk and CD ROM drives are disabled in the BIOS, and access to the BIOS protected by password.

Using an NT system policy the allowed software for each unique user ID is controlled. The allowed software consists of the synoptic control for that specific experimental area, and a means of viewing the status of any other named parameter on the control system. To prevent users from changing the system configuration, the desktop context menus (mouse right click), access to the registry, and access to Explorer, are all disabled.

## 2.8 System Management

Once the accounts are set up and working, there is little on-going management to perform. New accounts are created as experimental stations come on-line, and a database of machines and user IDs maintained.

Software updates required on the local hard disk (for example the ActiveX control) are done by visiting the PC and logging on locally, rather than updating over the network. This means that the use of the PC as an active control station is not compromised by remote administration.

## 3 COMMISSIONING

Installation and commissioning of the system had to be completed without any interruption to routine operation of the SRS and its experimental stations. This was achieved by adopting a staged programme of work. Initially, a simple beamline with only 3 experimental stations was transferred to the new system. This allowed us to gain experience in using and managing the new software and hardware before attempting upgrade of further stations. We also became familiar with the likely problems that may occur during commissioning. The remaining 30 stations were upgraded in 3 stages over a period of 12 months.

## 4 FUTURE DEVELOPMENTS

There is an increasing interest among the beamline users for a facility to integrate beamline control into the experimental control and data acquisition systems. This will probably be provided in the near future by implementing a gateway server with a simple socket based interfaced. This will use an ASCII protocol for sending and returning data to and from the control system. A security database will limit access to devices based on the IP address of the requesting host and the parameter being accessed. This facility will allow much greater automation of the data gathering process and simplification of the beamline user's task.

Also, there are plans to improve the online help system. At present, only basic operating instructions are available but complete beamline specific information, fault diagnosis instructions and operating procedures could easily be added.

## 5 CONCLUSIONS

The upgrade of the photon beamline control system was one of the last steps in replacing obsolete equipment in the SRS control system. It has now been completed and has brought about a significant improvement in control facilities on the experimental stations. All the changes were implemented during routine shutdown periods and no disruption to normal operating time occurred.

Some of the developments undertaken for the project, notably the Linux FEC design and the ActiveX control have wider use in the SRS control system and will prove useful in future projects.